\documentclass[%
    aip,
    jcp,
    amsmath,
    amssymb,
    twocolumn,
    superscriptaddress,
    reprint]{revtex4-1}
\usepackage{dcolumn}
\usepackage{graphicx} 
\usepackage{bm}
\usepackage{longtable}
\usepackage[english]{babel}
\usepackage{color}

\begin{document}

\title{Dielectric constant of water in the interface} 
\author{Mohammadhasan Dinpajooh} 
\author{Dmitry V.\ Matyushov}
\email{dmitrym@asu.edu}
\affiliation{Department of Physics and School of Molecular Sciences, 
         Arizona State University, PO Box 871504, Tempe, Arizona 85287}

\begin{abstract}
We define the dielectric constant (susceptibility) that should enter the Maxwell boundary value problem when applied to microscopic dielectric interfaces polarized by external fields. The dielectric constant (susceptibility) of the interface is defined by exact linear-response equations involving correlations of statistically fluctuating interface polarization and the Coulomb interaction energy of external charges with the dielectric. The theory is applied to the interface between water and spherical solutes of altering size studied by molecular dynamics (MD) simulations. The effective dielectric constant of interfacial water is found to be significantly lower than its bulk value, and it also depends on the solute size. For TIP3P water used in MD simulations the interface dielectric constant changes from 9 to 4 when the effective solute radius is increased from $\sim 5$ to 18 \AA.        
\end{abstract}

\maketitle 

\section{Introduction}
\label{sec:1}
The Maxwell field has played a prominent role in the theories of dielectrics for two mostly disconnected reasons. First, in the case of a homogeneous field produced by a planar capacitor, one gets the direct experimental access to the Maxwell field $E$ through the voltage on the plates $V$ and the distance between them $d$:  $E=V/d$. The response of the dielectric to  the external field is therefore conveniently represented in terms of the susceptibility to the Maxwell field. The second reason is that the scalar electrostatic potential $\phi$ is the solution of the dielectric boundary value problem when local constitutive relations are applied. The field $\mathbf{E}$ follows directly from that solution as $\mathbf{E}=-\nabla\phi$.\cite{Jackson:99} 

The first reason for the importance of $E$ can be viewed as both an advantage and disadvantage since $\mathbf{E}$ itself is never accessible experimentally and only the line integral $\int \mathbf{E}\cdot d\bm{\ell} = V$, producing the voltage $V$, can be measured.\cite{EygesBook:72} In the case of an inhomogeneous field there is no way to extract the field from the integral and experiment generally does not have direct access to the  inhomogeneous Maxwell field. The problem was realized already at the time of birth of the electromagnetic theory. Since inhomogeneous fields cannot be accessed directly, Thompson suggested to use small cavities to measure internal fields inside dielectrics.\cite{Thompson1872} This approach has in fact been realized by modern-day spectroscopy, which allows one to evaluate the local field acting on a dye molecule through the field-induced shift of its spectral line.\cite{Liptay:65,Fried:2015fh} However, the connection between such a local field and the macroscopic Maxwell field has been elusive beyond the standard prescriptions of the dielectric theory.\cite{DMjcp2:11} In addition, the ability to spatially resolve the distribution of the electric field and inhomogeneous polarization within molecular systems of nanometer scale has been limited.\cite{Steffen:94}

From the theoretical perspective, the Maxwell field is well defined by the Coulomb law. The starting point is the overall microscopic electric field $\mathbf{E}_m$, combining the field $\mathbf{E}_0$ of the external charges (vacuum field)  with the electric field of all molecular bound charges distributed with the charge density $\rho_b$ (``b'' stands for the bound charge). The result is obviously 
\begin{equation}
\mathbf{E}_m = \mathbf{E}_0 +\mathbf{E}_b,
\label{eq:1}  
\end{equation}
where 
\begin{equation}
\mathbf{E}_b= - \nabla \int \left|\mathbf{r}-\mathbf{r}'\right|^{-1} \rho_b' d\mathbf{r}' .
\label{eq:1-1}  
\end{equation}
The primes here and below denote vector and scalar fields taken at the point $\mathbf{r}'$, e.g., $\rho_b'=\rho_b(\mathbf{r}')$. The Maxwell field is produced from this equation as the result of two steps: (i) statistical average $\langle \mathbf{E}\rangle$ of the instantaneous $\mathbf{E}_m$ over the configurations of a statistical ensemble and (ii) coarse graining of $\langle \mathbf{E}\rangle$ over a ``physically small'' volume averaging out the microscopic correlations between the molecules of the material.\cite{Landau8} This volume is not precisely defined and, in fact, is never explicitly involved. The theory, as it is formulated for bulk dielectrics and interfaces, instead introduces coarse graining through constitutive relations as we discuss next.    

By taking the divergence of $\mathbf{E}_m$ and substituting $\nabla\cdot\mathbf{E}_0=4\pi\rho_0$ for the density of the external charge $\rho_0$, one arrives at $\nabla\cdot \mathbf{E}_m = 4\pi (\rho_0 + \rho_b)$. Further, due to the conservation of charge, the instantaneous density of bound charge can be replaced with the divergence of the polarization vector field $\mathbf{P}_m$, such as $\rho_b = -\nabla\cdot \mathbf{P}_m$.\cite{Landau8} One arrives at the equation for instantaneous fields
\begin{equation}
\nabla \cdot \left(\mathbf{E}_m + 4\pi \mathbf{P}_m\right) = 4\pi\rho_0 , 
\label{eq:2}  
\end{equation}
which looks very much like the standard Maxwell equation, except that the fields in this equation refer to an arbitrary statistical configuration of the system. Of course, this equation is just a different form of the Coulomb law, which applies to microscopic dimensions and arbitrary configurations of charges. The two-step averaging and coarse graining procedure mentioned above will produce the average smoothed-out fields $\mathbf{E}$ and $\mathbf{P}$ and the corresponding electric displacement vector $\mathbf{D}=\mathbf{E} + 4\pi\mathbf{P}$. The Maxwell equation for this coarse grained displacement vector follows from Eq.\ \eqref{eq:2} as $\nabla\cdot \mathbf{D}=4\pi\rho_0$.  

Equation \eqref{eq:2} and its coarse grained version still cannot be solved without applying a closure relation between $\mathbf{P}_m$ and $\mathbf{E}_m$ or between $\mathbf{P}$ and $\mathbf{E}$. The connection between microscopic fields $\mathbf{P}_m$ and $\mathbf{E}_m$ is a complex problem of statistical mechanics of liquids.\cite{Hansen:03} It is therefore assumed that coarse graining helps in eliminating this complexity and leads to local constitutive relations between coarse grained fields 
\begin{equation}
\mathbf{P}=\chi\mathbf{E} .
\label{eq:3}
\end{equation}
This constitutive relation thus establishes the direct proportionality between the vector fields $\mathbf{P}$ and $\mathbf{E}$ through the susceptibility $\chi$, which is a scalar for isotropic materials. Empirical evidence suggests that this approximation, when used for macroscopic dielectrics, yields the bulk dielectric susceptibility $\chi_s$, which is a material property, i.e., a parameter characterizing bulk dielectric and independent of the sample shape (the surface effects die off in the macroscopic limit). Correspondingly, the dielectric constant of bulk dielectric $\epsilon_s=1+4\pi\chi_s$ is a material property as well. 
   
This result is quite non-trivial since even for coarse grained vector fields the susceptibility $\chi_0$ to the field of external charges $\mathbf{E}_0$ does not share insensitivity to the surface effects (boundary conditions). $\chi_0$ is not a material property, and it depends on the shape of the sample through the dielectric boundary value problem. Given that the inhomogeneous Maxwell field $\mathbf{E}$ is not accessible experimentally, most problems of interest for applications involving inhomogeneous fields (solvation of molecules, solvent-induced shifts of spectral lines, interfacial problems, etc.) are formulated in terms of the response to an inhomogeneous external electric field $\mathbf{E}_0$. Nevertheless, the Maxwell field has to be introduced in order to solve the problem since only this field is believed to provide local constitutive relations between $\mathbf{E}$ and $\mathbf{P}$ required to arrive at the Poisson equation. The locality of the Maxwell field for inhomogeneous external fields does not have firm experimental support and is likely to be an approximation. This difficulty is responsible for many problems arising in the general problem of electric polarization of interfaces.\cite{DMjcp3:14,Jeanmairet:2013gl} 
 
The problem of interfacial polarization is solved in dielectric theories by replacing the microscopic fields $\mathbf{E}_m$ and $\mathbf{P}_m$ in each point of the interface with the corresponding coarse grained fields and then applying the local constitutive relation \eqref{eq:3} to each point of the interface.  When substituted to Eq.\ \eqref{eq:2}, it leads to the Poisson equation for $\mathbf{E}$ fully specified in terms of external charges. However, there is no factual coarse graining when this procedure is applied to microscopic problems, and it is nearly impossible even to define an algorithm of volume coarse graining when fields are changing on the scale of molecular dimensions. The Poisson equation is obtained in such cases by direct substitution $\mathbf{E}_m\to \mathbf{E}$ and $\mathbf{P}_m\to \mathbf{P}$ and the subsequent use of the constitutive equation. As mentioned, coarse graining of microscopic fields is not achieved directly by averaging over a judiciously chosen volume, but is produced by applying a specific local form of the constitutive relation. The smooth function $\mathbf{E}$ obtained from the solution of the Poisson equation then leads to a smooth $\mathbf{P}$, instead of a highly oscillatory function characteristic of interfaces.\cite{Ballenegger:05,Horvath:2013fe} It is the constitutive relation that replaces coarse graining over a small volume in converting the microscopic into macroscopic fields.

Since coarse graining is in fact not performed, one can adopt a somewhat different form of the constitutive relation involving only the statistically averaged fields in the interface
\begin{equation}
\langle \mathbf{P} \rangle = \chi\langle \mathbf{E} \rangle . 
\label{eq:4}  
\end{equation}
Of course, Eq.\ \eqref{eq:4} is an approximation. The question we address here is how to build a consistent theory of interfacial polarization when this approximation is applied. The advantage of Eq.\ \eqref{eq:4} over Eq.\ \eqref{eq:3} is that, in contrast to volume coarse graining, statistical averages are well-defined even on the microscopic length-scale and one can proceed with ensemble-based algorithms of defining susceptibilities. In other words, in contrast to smoothly varied functions $\mathbf{P}$ and $\mathbf{E}$ in Eq.\ \eqref{eq:3} the corresponding fields in Eq.\ \eqref{eq:4} will be highly oscillatory, as usually produced by liquid-state theories and numerical simulations. We do not intend to apply Eq.\ \eqref{eq:4} to the entire interface, but only to the dividing surface separating the solute from the solvent. In this way we connect the microscopic calculations to the electrostatic boundary value problem.   

If the constitutive relation is the only step separating the microscopic Coulomb law in Eq.\ \eqref{eq:2} from the dielectric boundary value problem, one wonders if this procedure can be supplemented with susceptibilities reflecting the microscopic structure of the polarized interface, i.e., the susceptibility $\chi$ in Eq.\ \eqref{eq:4}. The standard Maxwell dielectric boundary value problem in fact implements one additional approximation of replacing $\chi$ in Eq.\ \eqref{eq:4} with the susceptibility $\chi_s$ of bulk dielectric.\cite{Jackson:99} This approximation is not required and any scalar susceptibility can be used in solving the boundary value problem. Not surprising, the idea of an effective susceptibility or interfacial dielectric constant has been actively discussed in the literature.\cite{Ballenegger:03,Stern:2003cn,Ballenegger:05,Hansen:06,Bonthuis:2013fk}

 Some of phenomenological recipes proposed to deal with microscopic interfaces, such as the popular distance-dependent dielectric constant for solvation problems,\cite{Noyes:1962tj} do not withstand the scrutiny of microscopic formulations.\cite{DMjcp2:15} The problem with such formulations is that spatial correlation functions describing microscopic interfacial polarization are typically highly oscillatory\cite{Ballenegger:05,Horvath:2013fe} and do not allow defining simple distance-dependent susceptibilities. If any meaningful microscopic susceptibility has a chance to enter the standard  boundary value problem, it should be consistently derived from the microscopic Coulomb law in Eq.\ \eqref{eq:2} and not introduced as an \textit{ad hoc} phenomenological recipe justified by fitting to experimental data or results of numerical simulations. Providing such a consistent approach is the goal of this article. In other words, the main question addressed here is what is the dielectric constant, absorbing into itself the microscopic properties of the interface, that should enter the standard dielectric boundary value problem? We provide a general formulation of the problem, followed by specific calculations of the dielectric response of water interfacing a spherical solute.

\section{Boundary value problem}
\label{sec:2}
When one takes the statistical average in Eq. \eqref{eq:2}, one arrives at $\nabla\cdot \langle\mathbf{D}\rangle =0$ inside the dielectric where there are no external charges. This relation translates, through Gauss' theorem, into the condition of continuity of the projection of $\langle\mathbf{D}\rangle$ on the unit vector $\mathbf{\hat n}$ normal to the interface. This condition can be written as\cite{Landau8}
\begin{equation}
\mathbf{\hat n}\cdot\left(\nabla \langle \phi_1\rangle - \nabla\langle \phi_2\rangle \right) = 4\pi \sigma  , 
\label{eq:5}  
\end{equation}
where $\sigma$ is the surface charge density determined by the normal projections of the polarization density $P_{ni}=\mathbf{\hat n}\cdot\mathbf{P}_i$ ($i=1,2$) in two media in contact in the interface
\begin{equation}
\sigma = P_{n1}- P_{n2} . 
\label{eq:6}  
\end{equation} 

\begin{figure}
\includegraphics*[width=5cm]{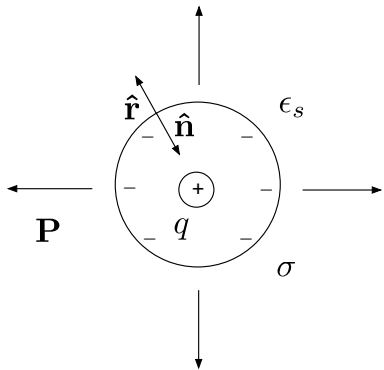}
\caption{Surface charge density at the interface between a spherical cavity and a dielectric with the dielectric constant $\epsilon_s$. The polarization density field $\mathbf{P}$ is aligned with the radial field of a positive charge $q$ placed at the center of the cavity. The surface charge density $\sigma$ is opposite in charge to $q$ to screen its interactions with charges placed outside of the cavity. $\mathbf{\hat n}$ denotes normal to the interface and $\mathbf{\hat r}=-\mathbf{\hat n}$ is the unit radial vector.   }
\label{fig:1}  
\end{figure}

In standard dielectric theories, the surface charge density $\sigma$ screens the external charge. It means that if a probe charge is placed at a large distance from the interface, the effective force between the external charge $q$ and the probe charge is reduced by the opposite charge of the interface polarization and an effective charge $q_\text{eff}$, instead of $q$, is measured by the force. This is illustrated in Fig.\ \ref{fig:1} for the simple case of a spherical cavity of radius $a$ with charge $q$ placed at its center, as discussed in the numerical simulations of aqueous solutions below. The positive charge $q$ will, in dielectric theories, create the opposite in sign surface charge density $\sigma = -(1-\epsilon_s^{-1})(q/S)$, where $S=\pi a^2$ is the area of the cavity. The effective charge producing the measurable force on an external probe charge, $q_\text{eff} =q +\sigma S=q/\epsilon_s$, is then reduced by the dielectric constant of the dielectric $\epsilon_s$.   

When the constitutive relation \eqref{eq:4} is applied to statistically averaged fields $\langle \mathbf{E}\rangle$ and  $\langle \mathbf{P}\rangle$ in Eq.\ \eqref{eq:2}, $\langle \mathbf{E}\rangle = -\nabla \langle \phi\rangle$ satisfies the Laplace equation $\Delta \langle \phi\rangle =0 $ inside the dielectric where there are no external charges. The properties of the interface enter the problem through the boundary condition in Eq.\ \eqref{eq:6}. Therefore, the goal of reformulating the standard Maxwell boundary value problem needs to focus on introducing microscopic properties of the interface into the boundary conditions of the Laplace equation. This is the goal we are pursuing here. 

Equation \eqref{eq:6} suggests that the only property of the interface one needs to supply to the solution of the Laplace equation is the surface charge density or the normal projection of the polarization density. The linear response approximation\cite{Hansen:03} provides the desired property in terms of a non-local susceptibility function $\bm{\chi}_0(\mathbf{r},\mathbf{r}')$ (generally a tensor) depending on two coordinates in the interface\cite{Felderhof:1977do,DMjcp1:04}  
\begin{equation}
\langle P_n(\mathbf{r})\rangle  = \int\mathbf{\hat n}\cdot\bm{\chi}_0(\mathbf{r},\mathbf{r}')\cdot \mathbf{E}_0'd\mathbf{r}' ,
\label{eq:7}
\end{equation} 
where the integral is over the entire space and the 2-rank tensor of susceptibility is
\begin{equation}
\bm{\chi}_0(\mathbf{r},\mathbf{r}') = \beta \langle \delta \mathbf{P}(\mathbf{r})\delta\mathbf{P}(\mathbf{r}')\rangle  . 
\label{eq:8}  
\end{equation}
Here, $\beta=1/(k_\text{B}T)$ and $\delta \mathbf{P} = \mathbf{P} - \langle \mathbf{P}\rangle$.  

The susceptibility $\bm{\chi}_0$ in Eq.\ \eqref{eq:8} is a second rank tensor defined by the corresponding Cartesian components.\cite{Felderhof:1977do} For some geometries of the interface, it is convenient to consider specific projections of $\bm{\chi}_0$. For instance, for the planar interface, one defines parallel ($\mathbf{\hat n}_\parallel$) and perpendicular ($\mathbf{\hat n}_\perp$) projections\cite{Stern:2003cn,Ballenegger:05} as the scalar functions $\chi_\parallel = \mathbf{\hat n}_\parallel \cdot \bm{\chi}_0\cdot \mathbf{\hat n}_\parallel$ and $\chi_\perp = \mathbf{\hat n}_\perp \cdot \bm{\chi}_0\cdot \mathbf{\hat n}_\perp$. Similarly, for spherical solutes which we consider below, one can define the scalar projection on the radial direction,\cite{Ballenegger:05} $\chi_0^{rr}=\mathbf{\hat r}\cdot \bm{\chi}_0\cdot\mathbf{\hat r}$. Such definitions become less useful for interfaces of arbitrary shape. Using longitudinal and transverse symmetries of the polarization field provides a more general formulation. \cite{DMjcp1:04,DMjcp2:04,Jeanmairet:2013gl} Our goal here does not involve calculating distance-dependent projections of the susceptibility. We focus instead on the normal projection of the polarization density field in Eq.\ \eqref{eq:6}, taken at the dividing surface, which can be defined for an arbitrary interface.

The two-point tensor $\bm{\chi}_0(\mathbf{r},\mathbf{r}')$ depends on two positions, $\mathbf{r}$ and $\mathbf{r}'$, separately to reflect its interface character and the involvement of three-body solute-solvent-solvent correlations. This needs to be contrasted with the non-local susceptibility of bulk dielectrics depending only on $\mathbf{r}-\mathbf{r}'$. To simplify the problem, one can assume that the length of polarization correlations in the interface is much shorter than the characteristic dimension of the interfacial region and apply the local approximation neglecting such correlations altogether\cite{Hansen:06}
\begin{equation}
\bm{\chi}_0(\mathbf{r},\mathbf{r}') = \delta (\mathbf{r}-\mathbf{r}')\bm{\chi}_0(\mathbf{r}') . 
\label{eq:9}  
\end{equation}
This approximation obviously eliminates the integral in Eq.\ \eqref{eq:7} shifting the focus to the inhomogeneous susceptibility $\bm{\chi}_0(\mathbf{r})$. It can be obtained by integrating Eq.\ \eqref{eq:8} over $\mathbf{r}'$
\begin{equation}
\bm{\chi}_0(\mathbf{r})=\beta \langle \delta \mathbf{P}(\mathbf{r})\delta\mathbf{M}\rangle,
\label{eq:10}  
\end{equation}
where $\mathbf{M}$ is the total dipole moment of the dielectric, $\delta\mathbf{M}=\mathbf{M} - \langle\mathbf{M}\rangle$. Analogs of this equation for different symmetries of the interface have been proposed by Stern and Feller\cite{Stern:2003cn} and by Ballenegger and Hansen\cite{Ballenegger:03,Ballenegger:05,Hansen:06} and extensively used in a number of recent simulations of interfacial polarization.\cite{Bonthuis:2011dq,Ghoufi:2012ad,Bonthuis:2013fk,Renou:2015jo,Schaaf:2015dl} We note that the local approximation becomes exact in the limit of a uniform external field considered by Stern and Feller.\cite{Stern:2003cn}

Before we proceed to the exact formula for the susceptibility tensor, not involving the local approximation of Eq.\ \eqref{eq:9}, it is useful to provide the connection between Eq.\ \eqref{eq:10} and the dielectric experiment performed by applying a uniform electric field to the bulk dielectric. One obtains for an isotropic dielectric
\begin{equation}
(\beta/\Omega)\langle (\delta\mathbf{M})^2\rangle = \Omega^{-1}\int_\Omega \mathrm{Tr}[\chi_0(\mathbf{r})] d\mathbf{r},
\label{eq:10-1}  
\end{equation}
where the integration is performed over the volume $\Omega$ of the dielectric and $\mathrm{Tr}[\bm{\chi}_0]=\sum_{\alpha}\chi_0^{\alpha\alpha}$. The fluctuation expression on the left-hand side of this equations enters the Kirkwood-Onsager equation for the dielectric constant\cite{Boettcher:73} and thus provides the connection between the volume integrated susceptibility to the bulk dielectric constant $\epsilon_s$. Such a connection is, however, not straightforward when one considers the distance dependence of a specific projection of $\bm{\chi}_0(\mathbf{r})$. In other words, polarization fluctuations $\delta \mathbf{P}(\mathbf{r})$ still carry microscopic information, no matter how far from the interface. These  fluctuations are coarse-grained, with the microscopic information lost, by volume integration. 

Even though the local approximation in Eq.\ \eqref{eq:9} provides a simple resolution of the problem, it is not required.\cite{DMjcp3:14} One can take into account the longitudinal character of the field of external charges $\mathbf{E}_0$ and the fact that $\int \mathbf{P}'_T \cdot \mathbf{E}_0' d\mathbf{r}'=0$ for the transverse projection of the polarization field $\mathbf{P}_T$ (Helmholtz theorem\cite{EygesBook:72}). Therefore, only the longitudinal projection of the polarization $\mathbf{P}'=\mathbf{P}'_L$ enters $\bm{\chi}_0$ in Eqs.\ \eqref{eq:7} and \eqref{eq:8}. The longitudinal projection of the polarization is in turn connected to the electrostatic field of the bound charge as\cite{DMjcp3:14}  $4\pi\mathbf{P}_L= - \mathbf{E}_b$, where $\mathbf{E}_b$ is given by Eq.\ \eqref{eq:1-1}. Given that $\mathbf{E}_b = -\nabla \phi_b$, one can apply the Gauss theorem to eliminate the integral in Eq.\ \eqref{eq:7}. The result is the exact relation for $P_n$ not requiring the use of the local approximation
\begin{equation}
\langle P_n\rangle  = - \beta \langle \delta P_n \delta U^C\rangle .
\label{eq:11}  
\end{equation}
Here, 
\begin{equation}
\delta U^C = \sum_i q_i \delta  \phi_{bi}
\label{eq:12}  
\end{equation}
is the fluctuation of the Coulomb interaction energy of the dielectric with the external charges $q_i$; $\phi_{bi}$ is the electrostatic potential of the bound charges of the dielectric at the location of the charge $q_i$, $\delta \phi_{bi} =\phi_{bi} - \langle \phi_{bi}\rangle$.   

It is important to emphasize that no specific assumptions regarding either the origin of the polarization density $P_n$ or the electrostatic energy $U^C$ have been introduced in deriving Eq.\ \eqref{eq:11}. Both parameters can be microscopic quantities sampled by numerical computer simulations. For instance, a polar liquid with molecular dipoles $\mathbf{m}_j$ with coordinates $\mathbf{r}_j$ will have the polarization density $P_n(\mathbf{r})=\mathbf{\hat n}\cdot\sum_j\mathbf{m}_j\delta\left(\mathbf{r}-\mathbf{r}_j\right)$. Correspondingly, $U^C$ can be viewed as the energy of Coulomb interactions of all atomic partial charges of the dielectric with the external charges. This property is routinely provided by numerical simulations. 

It is also important to emphasize that although the polarization density projection $P_n$ is calculated here for the interface and thus can be viewed as a local property, the non-local character of the Coulomb interactions of bound charges are incorporated into $U^C$. The correlation function in Eq.\ \eqref{eq:11} is fundamentally a three-body, solute-solvent-solvent correlation function, incorporating all chain diagrams of dipolar interactions responsible for dielectric screening.\cite{Wertheim:71}   

Equation \eqref{eq:11} is the exact solution for the problem of the surface charge density in the interface assuming linear response to the field of external charges. Deriving it does not require constitutive relations. If the constitutive relation \eqref{eq:4} is adopted, one can find the statistically averaged electrostatic potential of the interface $\langle \phi\rangle$ from the solution of the Laplace equation and, in addition, ask the question of what susceptibility, or dielectric constant, can be assigned to the interface. Such scalar interface susceptibility can be defined by the equation 
\begin{equation}
  \langle P_n\rangle = \chi_{0n} E_{0n} .   
  \label{eq:13}
\end{equation}
This constitutive relation, even though it looks similar to Eq.\ \eqref{eq:3}, is in fact a weaker condition. It applies to one projection only, instead to all three Cartesian projections in Eq.\ \eqref{eq:3}, and only in the interface, instead of the entire bulk of the dielectric as assumed in Eq.\ \eqref{eq:3}.

\begin{figure}
\includegraphics*[width=7cm]{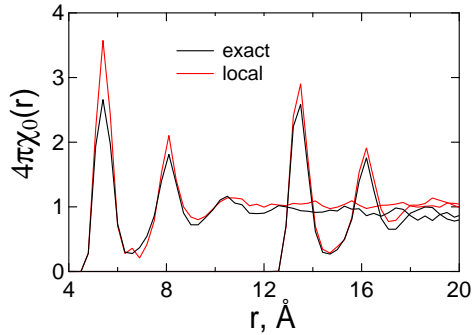}
\caption{Comparison of the local [Eq.\ \eqref{eq:14}] and exact [Eq.\ \eqref{eq:15}] formulas for the interface susceptibility of TIP3P water interfacing Kihara solutes of different size. The Kihara solutes are characterized by the hard-sphere core of the radii $R_\text{HS}=2$ and 10 \AA\ and the Lennard-Jones (LJ) diameter of $\sigma_\text{LJ} =3$ \AA\ for the LJ interaction between the solute and water's oxygen. The position of the first peak of the radial distribution function is approximately located at $R_\text{HS}+\sigma_\text{LJ}$. 
}
\label{fig:2}  
\end{figure}

Equation \eqref{eq:13} can be substituted back to Eqs.\ \eqref{eq:5} and \eqref{eq:6} to produce the boundary conditions for the Poisson equation
\begin{equation}
\mathbf{\hat n}\cdot\left(\nabla \langle \phi_1\rangle - \nabla\langle \phi_2\rangle \right) = 4\pi \left(\chi_{0n,2} -\chi_{0n,1}\right)\mathbf{\hat n}\cdot \nabla \phi_0,
\label{eq:13-1}  
\end{equation}
where $\phi_0$ is the electrostatic potential of external charges remaining continuous at the dividing surface. This is the only place where the susceptibility of the interface enters the boundary value problem. The local constitutive relations, Eqs.\ \eqref{eq:3} and \eqref{eq:4}, applied globally to the entire dielectric sample in dielectric theories, are replaced with the constitutive relation in Eq.\ \eqref{eq:13} applied to the dividing surface only.

The constitutive equation \eqref{eq:13} might be a reasonable approximation for a few molecular layers in the interface, but is not expected to hold globally. Likewise, the susceptibility $\chi_{0n}$, and the interface dielectric constant $\epsilon_\text{int}$ defined for spherical solutes below, are parameters characterizing the interface. We, therefore, do not expect them to approach the dielectric susceptibility or the dielectric constant of the bulk material in any specific limit. Even for a macroscopic interface, $\chi_{0n}$ is still an interfacial parameter (like the surface tension), which should not be expected to be simply related to the bulk susceptibility $\chi_s$.

\section{Interface of a spherical solute}
\label{sec:3}     
Here we apply the arguments presented above to the problem of water polarization at the interface of a spherical solute. The normal to the interface is defined outward from the dielectric,\cite{Landau8} $\mathbf{\hat n}=-\mathbf{\hat r}$, $\mathbf{\hat r}=\mathbf{r}/r$ (Fig.\ \ref{fig:1}). A further simplification of the geometry is achieved by locating the external charges at the center of the solute. All susceptibility tensors  become scalars, with the only non-zero diagonal radial component $\chi_0^{rr}=\mathbf{\hat r}\cdot \bm{\chi}_0\cdot\mathbf{\hat r}$. We will drop the indexes for brevity with the notation $\chi_0(r) = \chi^{rr}_0(r)$. For this specific type of interface, the local approximation leads to
\begin{equation}
\chi_0(r) = \beta \langle \delta P_r(r)\delta M_r\rangle,
\label{eq:14}  
\end{equation}
where $P_r=-P_n$ and $M_r=-M_n$ denote the radial projections of the corresponding vectors (Fig.\ \ref{fig:1}). 

Since the electric field of the central charge $q$ is $E_{0n}=-q/r^2$, one can define the linear distance-dependent susceptibility $\chi_{0n}(r)$ analogous to the one in Eq.\ \eqref{eq:13}
\begin{equation}
\chi_{0n}(r) = - \beta r^2 \langle \delta P_r(r)\delta \phi_b\rangle .
\label{eq:15}  
\end{equation}
Here, $\phi_b$ is the electrostatic potential produced by the dielectric at the center of the spherical solute where the external charge is placed. The interface susceptibility $\chi_{0n}$ follows from this function by adopting $r=a$, i.e., the radius of the spherical surface separating the solute from the surrounding dielectric. We note that our Eq.\ \eqref{eq:15} is equivalent to its integral form earlier derived by Ballenegger and Hansen\cite{Ballenegger:05}
\begin{equation}
\chi_{0n}(r) = 4\pi r^2\beta\int_a^{\infty}  \langle \delta P_r(r)\delta P_r(r')\rangle dr',
  \label{eq:15-2}
\end{equation}
where the integral over $r'$ produces $\delta \phi_b$ in our Eq.\ \eqref{eq:15}.

The definition of the position of the dividing dielectric surface presents a major difficulty for all dielectric theories, and it is not going to go away in our formulation recasting the problem of a microscopic polarized interface as the dielectric boundary problem. The question we are addressing here is what is the susceptibility or the surface charge density that needs to enter the boundary conditions once such a dividing surface is defined. A question relevant to this goal is how sensitive such a definition would be to possible variations of the position of the dividing surface. One ideally wants a robust definition, little sensitive to changes in the cavity radius $a$.

\begin{figure}
\includegraphics*[width=7cm]{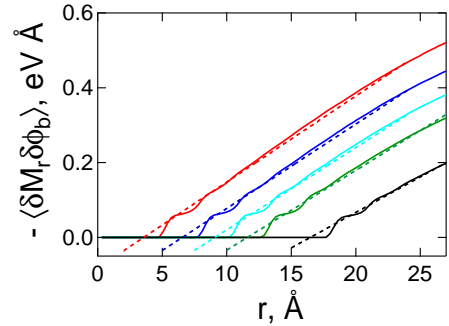}
\caption{Definition of the interface susceptibility $\chi_{0n}$ in terms of the slope of $-\langle \delta M_r(r) \delta \phi_b\rangle$ according to Eq.\ \eqref{eq:16}. The dashed lines show linear fits to $-\langle \delta M_r(r) \delta \phi_b\rangle$ calculated from MD simulations of Kihara solutes interfacing TIP3P water. The hard-sphere core of the Kihara solutes was varied in simulations: $R_\text{HS}=2$ (red), 5 (blue), 7.5 (cyan), 10 (green), and 15 (black) \AA. }   
\label{fig:3}  
\end{figure}

Figure \ref{fig:2} shows $\chi_0(r)$ in the local approximation [Eq.\ \eqref{eq:14}] and the exact $\chi_{0n}(r)$  [Eq.\ \eqref{eq:15}] calculated from molecular dynamics (MD) simulations performed in this study. The simulations are done for TIP3P water\cite{tip3p:83} interfacing spherical solutes of varying diameter and interacting with the oxygen of water by the Kihara potential (a hard-sphere repulsion core with the radius $R_\text{HS}$ combined with a surface layer of soft Lennard-Jones potential).\cite{Kihara:58,DMcpl:11} One has to keep in mind that correlation functions in Eqs.\ \eqref{eq:14} and \eqref{eq:15} fundamentally reflect three-particle solute-solvent-solvent correlations. Relatively long MD simulations, $\sim 200$ ns, were therefore required to converge them for each solute studied here. More detail on the simulation protocol is given in the supplementary material,\cite{supmatJCP} and here we discuss the results. 

It is clear from the calculations that the local [Eq.\ \eqref{eq:14}] and exact [Eq.\ \eqref{eq:15}]  formulations for the radial interface susceptibility generally agree with each other. The exact formulation is obviously preferable since it is free of the locality assumption. Both results show an oscillatory behavior of the interface susceptibility, leading to potential uncertainties when the cavity radius is altered. Some type of averaging over the oscillations, or coarse graining, is needed to arrive at a robust definition of interface susceptibility and the corresponding dielectric constant. An approach developed previously\cite{DMjcp3:14} and adopted here is to represent $\chi_{0n}(r)$ in Eq.\ \eqref{eq:15} as the derivative of the correlation function based on the integrated dipole moment $M_r(r)$ of water within the sphere of radius $r$ 
\begin{equation}
\chi_{0n}(r) = -\frac{\beta}{4\pi}\frac{d}{dr}\langle \delta M_r(r) \delta \phi_b\rangle . 
\label{eq:16}  
\end{equation}
If differential in the above equation is taken at each point, one recovers $\chi_{0n}(r)$, with its oscillatory behavior shown in Fig.\ \ref{fig:2}. Alternatively, instead of taking the differential at each point, we determine the linear slope in respect to $r$ to average out the oscillations of $\chi_{0n}(r)$ caused by molecular granularity. This linear slope then provides us with the scalar coarse grained susceptibility of the interface with oscillations averaged out. Figure \ref{fig:3} shows that indeed the slope can be well defined from the correlation function $\langle \delta M_r(r) \delta \phi_b\rangle$ calculated in a few molecular layers. 

\begin{figure}[t]
\includegraphics*[clip=true,trim= 0cm 1.3cm 0cm 0cm,width=7cm]{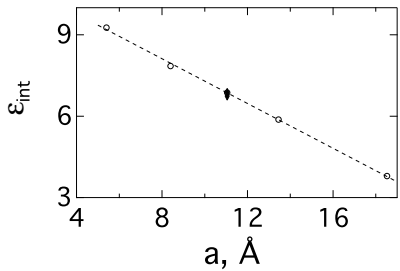}\\

\caption{Interface dielectric constant $\epsilon_\text{int}$ plotted against the cavity radius $a=r_\text{max}$ defined as the distance $r_\text{max}$ to the first peak of the solute-oxygen pair distribution function. Circles refer to neutral Kihara solutes, while diamonds refer to anion and cation Kihara solutes ($R_\text{HS}=10$ \AA, $r_\text{max}=11.05$ \AA) with charges $q=\pm 1$ placed at the solute's center (not distinguishable on the scale of the plot). The dashed line is a linear regression through the points drawn to guide the eye.    
}
\label{fig:4}   
\end{figure}

The susceptibility $\chi_{0n}$ to a radial external field $E_{0n}$ can be associated with the interface dielectric constant  $\epsilon_\text{int}$ according to the relation $(\epsilon_\text{int}-1)/(4\pi \epsilon_\text{int})=\chi_{0n}$,\cite{Ballenegger:05,DMjcp3:14} which leads to
\begin{equation}
  \epsilon_\text{int} = \left[1-4\pi\chi_{0n}\right]^{-1} . 
  \label{eq:17}
\end{equation}
The values, obtained from the slopes of the radial correlation functions shown in Fig.\ \ref{fig:3}, are presented in Fig.\ \ref{fig:4}. We find that $\epsilon_\text{int}$ decreases slowly from $\sim 9$ to $4$ as the effective size of the solute increases from $\sim 5.5$ \AA\ to $18.5$ \AA.  Overall, the value of the interface dielectric constant is much smaller than the  bulk value for TIP3P water, $\epsilon_s\simeq 97$.\cite{Guillot:02} While there is no \textit{a priori} reason to anticipate $\epsilon_\text{int}=\epsilon_s$, it is this assumption that is used in the standard dielectric boundary value problem.\cite{Jackson:99} We also note that the definition of $\epsilon_\text{int}$ by Eq.\ \eqref{eq:17} is prone to numerical instabilities when $4\pi\chi_{0n}$ becomes greater than unity due to calculation errors. $\epsilon_\text{int}$ is not required for the solution of the boundary value problem in Eq.\ \eqref{eq:13-1} and $\chi_{0n}$ is sufficient. It is presented here solely because of the history of the subject casting the dielectric boundary value problem in terms of the dielectric constant.

The results shown by circles in Fig.\ \ref{fig:4} are obtained for neutral Kihara solutes. Even though Eq.\ \eqref{eq:11} contains the electrostatic interaction energy with the external charge of the ion, which is proportional to the charge magnitude, charge cancels out when the surface susceptibility is defined by dividing the surface polarization by the ion field in Eq.\ \eqref{eq:13}. We therefore operate in the linear response domain, when one can assume that the presence of the external charge does not alter the structure of the interface used to perform the statistical averages. That this is indeed the case is demonstrated by simulating Kihara solutes with positive and negative charges placed at their centers. These results are shown by diamonds in Fig.\ \ref{fig:4} and are indistinguishable from the results obtained for neutral solutes (see supplementary material for more detail\cite{supmatJCP}). Our simulations are indeed consistent with the linear response approximation. 

The small value of the interface dielectric constant of water has potentially dramatic consequences for the problem of hydration. Our formalism anticipates that $\epsilon_\text{int}$ is used in the dielectric boundary value problem. Therefore, the solvation free energy of a spherical ion carrying charge $q$ and assigned the cavity radius $a$ is given by the standard Born equation\cite{Born:20,Hummer:96} $F = -(1/2)\chi_B q^2$, where the Born solvation susceptibility is
\begin{equation}
\chi_B =\frac{4\pi}{a}\chi_{0n}(a)= \frac{1}{a}\left( 1 -\frac{1}{\epsilon_\text{int}(a)}\right) .
\label{eq:18}  
\end{equation}
Here, $\chi_{0n}(a)$ and $\epsilon_\text{int}(a)$ indicate that the dependence of the Born solvation susceptibility on the cavity radius can be more complex than the traditionally anticipated $a^{-1}$ scaling. 
 
The reasons for the relative success of the Born equation in predicting the free energy of solvation and its dramatic failure in describing entropy of solvation have long been known.\cite{Roux:90,Rick:94,Lynden-Bell:1997uq,Rajamani:04} Both are related to the low sensitivity of the Born formula to the solvent properties when the bulk dielectric constant $\epsilon_s\gg 1$ is used instead of $\epsilon_\text{int}$ in Eq.\ \eqref{eq:18}. The traditional form of the Born equation significantly underestimates the entropy of solvation\cite{Roux:90} since the term $\epsilon_s^{-2}\partial \epsilon_s/\partial T$, appearing in the entropy, is too small. This deficiency can be potentially remedied if, according to our calculations, $\epsilon_\text{int}\ll \epsilon_s$. The final verdict  requires knowledge of $\epsilon_\text{int}(T)$. Our estimate for TIP3P water gives  $\left(\partial\epsilon_\text{int}/\partial T\right)_V\simeq -0.8\times 10^{-4}$ K$^{-1}$ (supplementary material\cite{supmatJCP}), which is significantly lower than $\left(\partial\epsilon_s/\partial T\right)_P =-0.36$ K$^{-1}$ of bulk water. Whether this low value is shared by more realistic force fields of water is not clear at the moment.

\section{Conclusions}         
\label{sec:4}
We discuss here a formalism connecting the Maxwell boundary value problem with the microscopic structure of the interface. In other words, the paper asks the question: What is the dielectric constant that should enter the boundary conditions in the Laplace equation describing a polarized dielectric interface? The problem is formulated in terms of the interface susceptibility or, alternatively, the interface dielectric constant. This property is calculated from an exact equation statistically averaging correlated fluctuations of the interface polarization density and the electrostatic energy of external charges interacting with the polarized dielectric. Evaluated by MD simulations of water interfacing spherical solutes, the interface dielectric constant is found to be significantly lower than the corresponding bulk value.

\acknowledgments 
This research was supported by the National Science Foundation (CHE-1464810) and through XSEDE (TG-MCB080116N).  The authors are grateful to Daniel Martin for his help with the setup of simulations. 


%

\end{document}